\documentclass[graybox]{svmult}

\usepackage{type1cm}        
\usepackage{makeidx}         
\usepackage{graphicx}        
\usepackage{multicol}        
\usepackage[bottom]{footmisc}

\usepackage{newtxtext}       %
\usepackage{newtxmath}       

\usepackage{colortbl}  

\definecolor{lightgray}{gray}{0.9}

\usepackage{array}
\usepackage{enumitem} 
\newlist{todolist}{itemize}{2}
\setlist[todolist]{label=$\square$}
\usepackage[most]{tcolorbox} 

\makeindex             

\begin{document}

\newtcolorbox{mybox}{
  colback=lightgray, 
  colframe=lightgray, 
  coltext=black, 
  boxsep=5pt, 
  arc=0pt, 
  breakable, 
}

\newtcolorbox{notebox}{
  colback=darkgray, 
  colframe=darkgray, 
  coltext=white, 
  boxsep=5pt, 
  arc=0pt, 
  breakable, 
}

\title*{Teaching Research Design in Software Engineering}
\author{Jefferson Seide Molléri, Kai Petersen}
\institute{Jefferson Seide Molléri \at Kristiania University College, Norway, \email{jefferson.mollleri@kristiania.no} 
\and Kai Petersen \at Flensburg University of Applied Sciences, Germany, \email{kai.petersen@hs-flensburg.de}\\
Blekinge Institute of Technology, Sweden, \email{kai.petersen@bth.se}}
%
%
\maketitle

\begin{notebox}
\textit{Note: This chapter is part of the upcoming book titled "Teaching Empirical Research Methods in Software Engineering," to be published by Springer.}
\end{notebox}

In the dynamic field of Software Engineering (SE), where practice is constantly evolving and adapting to new technologies, conducting research is a daunting quest. This poses a challenge for researchers: \textit{how to stay relevant and effective in their studies?} Empirical Software Engineering (ESE) has emerged as a contending force aiming to critically evaluate and provide knowledge that informs practice in adopting new technologies. Empirical research requires a rigorous process of collecting and analyzing data to obtain evidence-based findings. Challenges to this process are numerous, and many researchers, novice and experienced, found difficulties due to many complexities involved in designing their research. 

The core of this chapter is to teach foundational skills in research design, essential for educating software engineers and researchers in ESE. It focuses on developing a well-structured research design, which includes defining a clear area of investigation, formulating relevant research questions, and choosing appropriate methodologies. While the primary focus is on research design, this chapter also covers aspects of research scoping and selecting research methods. This approach prepares students to handle the complexities of the ever-changing technological landscape in SE, making it a critical component of their educational curriculum.

\begin{mybox}
    \textbf{Teaching Modules}. In this chapter, we incorporate active learning principles to enhance student engagement through practical activities and reflection. Following the constructive alignment framework \cite{biggs2022teaching}, we formulate teaching modules that include Intended Learning Outcomes (ILOs), Teaching-Learning Activities (TLAs), and suggested assessments. For ease of pedagogical application, these teaching modules are clearly highlighted at the end of each relevant subsection.
\end{mybox}

\section{Research Design at a Glance}
\label{sec:at_glance}

This chapter outlines a systematic approach to designing research in SE, combining structured methodology with practical advice to help researchers effectively navigate the complexities of SE research. Additionally, we offer guidance on teaching each step of the research design process. 

\begin{figure}[!ht]
\centering
    \includegraphics[width=\textwidth]{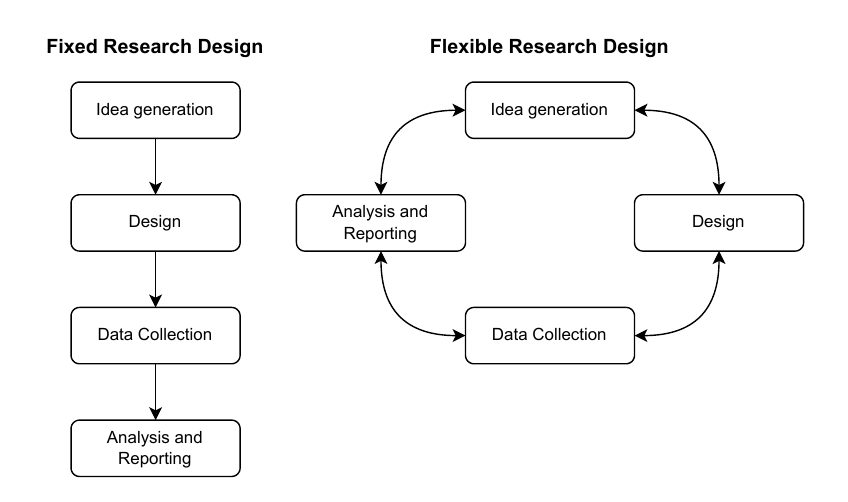}
    \caption{Procedures for (a) fixed and (b) flexible research designs, according to \cite{kampenes2008flexibility}.}
    \label{fig:fixedvsflexible}
\end{figure}

The approach to research design can vary significantly, generally categorized into two types, as shown in Figure \ref{fig:fixedvsflexible}. \textbf{Fixed designs} are highly structured and follow a systematic progression through clearly defined phases, typical for studies that utilize structured methodologies. This offers clarity and predictability, with each phase building directly on the one before \cite{robson2016real}. In contrast, \textbf{flexible designs} are iterative and adaptive, ideal for research projects where the scope isn't initially clear and may evolve with new findings or trends \cite{kampenes2008flexibility, robson2016real}. 

\begin{figure}[!ht]
\centering
    \includegraphics[width=.9\textwidth]{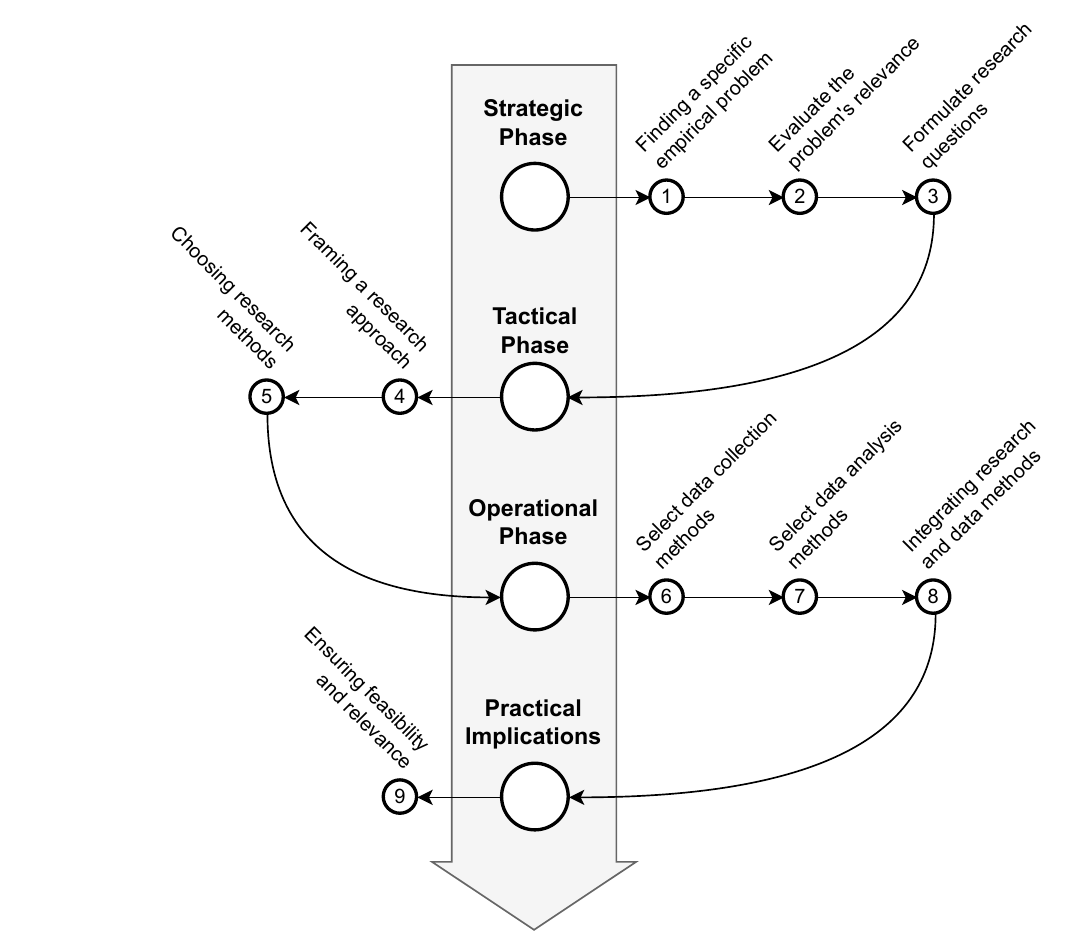}
    \caption{Overview of the research design process.}
    \label{fig:framework-fixed}
\end{figure}

The generic framework for research design depicted in Figure \ref{fig:framework-fixed} provides the overarching structure for this chapter. It involves nine methodological steps, covering research strategy formulation (steps 1-3), tactical decisions about the overall research method (steps 4 and 5), selection of operational methods (steps 6-8), and practical considerations (step 9).

While this framework is better suited for fixed research designs, it can also fit flexible designs with minor modifications. In flexible research design, the framework's steps are not required to be completed sequentially before data collection begins. Instead, multiple iterations may refine and adjust the research process as more data are collected and analyzed. Additionally, adjustments in the research design can also lead to changes in idea generation and other phases of the research project.

Educators should strive to teach both fixed and flexible design paradigms, providing students with the tools to choose the appropriate approach based on their research goals. Practically, researchers should consider their project's scope, the nature of the research question, and the degree of uncertainty in the data when deciding between a fixed and a flexible design.

\section{Finding a Relevant Empirical Problem}
\label{sec:empirical_problem}

Finding a relevant problem in SE requires understanding of the field's dynamic nature and how it shapes research directions. This section begins by exploring the evolving landscape of SE and its implications for research, then defines criteria for determining the relevance of a research problem.

\subsection{The Current State of SE Research and Practice}
\label{sec:current_state}

Software Engineering is a broad field that encompasses both academic study and practical applications. This duality, common to applied fields, offers unique advantages and challenges when identifying a potential problem that contributes to the body of knowledge. This knowledge spans the \textit{state of practice}, i.e., practical applications and standards known, accepted, and adopted by professionals, and the \textit{state of the art}, i.e., the recent findings and theories that have been attested by empirical research.

Integral to SE is its broad spectrum of knowledge areas, which include diverse topics ranging from software requirements and testing to software engineering management, each offering distinct perspectives and insights into the field. Many of the established knowledge areas related to SE are outlined by the Software Engineering Body of Knowledge \cite{bourque2014swebok}. 

Currently, SE research is on cutting-edge areas such as Artificial Intelligence (AI) and machine learning, cloud computing, DevOps, and cybersecurity. Agile methodologies, while still pivotal, are evolving to emphasize scalability and adaptability. In practice, there's a strong trend towards automating development processes and incorporating data-driven decision-making. Additionally, ethical considerations and sustainability in software development are gaining traction. This evolving landscape reflects the dynamic nature of SE and also underscores the need for research that is relevant to practice.

\subsection{Sources for Identifying Problems}

Essentially, a research problem is a specific issue or gap in the existing knowledge that research aims to address or resolve. Thus, a well-defined research problem typically goes hand-in-hand with an identified knowledge gap in the existing body of knowledge.

\subsubsection{Analyzing Existing Literature}

Analyzing the existing literature is a first step to identify obvious knowledge gaps. The body of knowledge represented by literature (i.e., state of the art) is a source for research problems of interest. Research articles serve as a primary source to identify problems the industry is currently facing, and outlining topics for further investigation. For example, a researcher interested in the challenges of integrating AI into software testing would start by examining the current literature on AI techniques within SE to determine what has been extensively studied and where gaps exist.

A thorough systematic literature review (SLR, see Chapter Y) offers a structured approach to survey the existing body of knowledge and obtain a comprehensive overview of a topic. However, researchers should consider the possible delay in academic literature reflecting the latest advancements, and thus, might also include grey literature like blogs and whitepapers for a more current perspective \cite{garousi2019guidelines}.

\subsubsection{Bridging Theory and Practice}

Researchers should extend their focus beyond academic context and actively build relationships with companies to integrate practical insights into their research design. The diverse practices and the multi-faceted aspects of the software process adopted by companies are key sources for identifying current problems in SE. Therefore, researchers should seek a dialogue with industry for formulating relevant research problems.

Engaging with practitioners that have extensive experience and understanding of a particular field helps identify under-researched areas and validate the most current state of practice. For example, software development teams using Agile methodologies can reveal adoption rates, challenges, benefits, and organizational impacts of these practices, offering a practical context that academic literature alone may not provide. 

Suggested steps for industry engagement:

\begin{itemize}
    \item Partner with companies through joint research and development (R\&D) projects to gather data on relevant phenomena and industry needs.
    \item Attend industry conferences and workshops to stay updated on current practices and emerging trends, and to consult with experts for practical insights.
    \item Share preliminary findings with industry partners and incorporate their feedback to refine research focus and ensure it addresses real-world problems.
\end{itemize}

Building on the previous example, after identifying gaps in integrating AI into software testing from the literature, researchers consulted industry experts to understand the practical challenges faced when adopting AI in testing environments. This led them to review project post-mortems from development teams that had attempted AI solutions, uncovering common issues such as integration complexity and tool usability. These insights helped define a specific empirical problem: the difficulties in integrating AI for automated test generation in agile development cycles.

\subsubsection{Validating Knowledge Gaps}

To ensure the integrity of identified research problems, researchers should cross-reference information from various sources, like literature reviews and expert consultations. Additionally, open-access databases and industrial publications help enhancing the trustworthiness and accuracy of the information obtained. Using this multi-faceted approach, researchers can confirm that the identified gaps are not only theoretically significant but practically relevant.

\begin{mybox}
    \textbf{Teaching Module 1}
    
    \textbf{Intended Learning Outcome:} Develop the ability to formulate a well-defined research problem in SE.
    
    \textbf{Teaching and Learning Activities:} Have students develop a research proposal, including a clearly defined problem statement. This task should involve background research, stating the significance of the problem, and hypothesizing potential outcomes. 

    After developing research proposals, facilitate peer-review sessions where students critique and provide feedback on each other’s problem statements, enhancing their ability to identify and articulate research problems.

    \textbf{Assessment:} Students will be assessed on the quality of their feedback and  their ability to critically analyze the research problems posed by peers.
\end{mybox}

\subsection{Criteria for Relevance}

Relevance refers to the degree to which research findings are applicable and beneficial to real-world problems and practices. The concept of relevance commonly adopted in SE is inherited from Information Systems (IS) research \cite{benbasat1999empirical}, in which qualities such as being of future interest of practitioners, applicable in the current state of practice, and written in a style that professionals would understand, are worthy of investigation. 

Petersen et al. \cite{petersen2024revisiting} further refine this by proposing a framework to assess industry relevance in SE research. This framework evaluates research contributions across six key aspects: What, How, Where, Who, When, and Why, each addressing different dimensions of a project. For example, when assessing the relevance of a new testing framework for continuous integration, researchers might evaluate its application in various industry settings (Where), its alignment with current testing practices (How), and its potential benefits for software quality (What). This multi-dimensional relevance assessment helps understanding the feasibility, applicability, and generalizability of research findings in SE.

\begin{mybox}
    \textbf{Teaching Module 2}
    
    \textbf{Intended Learning Outcome:} Develop the ability to critically reflect on the relevance of research problems in SE.

    \textbf{Teaching and Learning Activities:} Assign tasks where students must perform relevance assessment assignments and write about the relevance of a chosen research problem (e.g. from real case studies), including the use of literature on research relevance definition. 

    Organize debate sessions on the relevance and impact of various research problems in SE, helping students to articulate and defend their perspectives. 

    \textbf{Assessment:} Students must submit written assignments demonstrating their ability to articulate and defend their evaluations of research problem relevance
\end{mybox}

\section{Formulating Research Questions}
\label{sec:research-questions}

Research questions (RQs) set the direction of the investigation efforts and determine how research is conducted. They are also essential for the further steps of research design, i.e., choosing a research method and detailing a research approach. A good research question should be clear and focused, stating precisely what is intended to be studied and avoiding ambiguities. It should also addresses a significant gap in the existing body of knowledge, as stated in the previous step.

Consider the following poorly formulated RQ: \textit{"How can software development be improved?"} This question is too broad and lacks specificity. It does not outline a particular aspect to be investigated, the variables or the context involved. A more effective RQ could be: \textit{"How does continuous integration and deployment (CI/CD) practices impact the software release cycle time and bug detection rate in small-sized software development organizations?"} This question targets a specific topic within the SE field, outlines the variables to be studied, and also defines the context by means of a population.

Well-formulated research questions not only guide the study's direction but also shed light on the most suitable methodologies to use. For example, our example of a well-formulated RQ above suggests using a quantitative approach, specific measurements like release cycle intervals and bug detection rates, and a comparison of different practices, such as the implementation versus non-implementation of CI/CD practices.

\subsection{Crafting Effective Research Questions}

In selecting empirical methods for SE research, Easterbrook et al. \cite{easterbrook2008selecting} distinguished different types of research questions that drive the choice of research and data collection methods. Table \ref{tab:types-RQs} provides an overview of the question types.

\begin{table}[!ht]
\caption{Types of research questions, according to Easterbrook et al. \cite{easterbrook2008selecting}.}
\label{tab:types-RQs}
\rowcolors{1}{white}{lightgray}
\begin{tabular}{>{\raggedright}p{2cm}p{4.7cm}p{4.7cm}}
\hline\noalign{\smallskip}
\textbf{Type} & \textbf{Structure} & \textbf{Example} \\ 
\noalign{\smallskip}\svhline\noalign{\smallskip}
Existence & 
Does [phenomenon] exist in [context]? &
Does \textit{continuous integration} exist in \textit{small software development teams}? 
\\
Description and Classification & 
What are the [characteristics] of [context]? &
What are the \textit{common challenges} faced by \textit{teams} during the \textit{adoption of Agile methodologies}?
\\
Descriptive-Comparative & 
How does the [aspect] of [group 1] compare to [group 2]? &
How does the \textit{code quality} of \textit{projects using test-driven development} compare to \textit{those not using it}?
\\
Descriptive-Process & 
How is the [process] [action] in [context]? &
How is the \textit{deployment process} typically \textit{executed} in \textit{DevOps environments}?
\\
Relationship & 
Is there a [relationship] between [variable 1] and [variable 2] in [context]? &
Is there a \textit{correlation} between \textit{team size} and \textit{productivity} in \textit{software development projects}?
\\
Causality & 
Does [event] causes/prevents the [effect] in [context]? &
Does \textit{the use of automated testing} reduce \textit{the number of defects} in \textit{software releases}?
\\
Causality-Comparative &
Does [condition 1] cause/prevent [effect] than [condition 2] in [context]? &
Does \textit{pair programming} lead to \textit{fewer bugs} than \textit{solo programming} in \textit{small-scale software development}?
\\
Causality-Comparative Interaction & 
Does [variable 1] on [outcome] differ across [conditions]? &
Does the effect of \textit{continuous deployment} on \textit{release frequency} vary with \textit{team size}?
\\
Design & 
What is an effective way to [action] into the [context]? &
What is an effective way to \textit{integrate security practices} into \textit{the Agile development lifecycle}?
\\
\noalign{\smallskip}\hline\noalign{\smallskip}
\end{tabular}
\end{table}

From the questions it already becomes clear that they require different methods to be answered. Descriptive-Process questions focusing on detailed process descriptions require to gather deep insights of how processes are executed, and hence require detailed examinations of real processes with a variety of options (such as case studies, observations, or ethnographies). Causality questions require controlling confounding factors and hence point towards experimental studies. 

Wohlin and Arum \cite{wohlin2015towards} introduced various decision points to arrive at a research design. They considered the definition of the research outcome (basic vs. applied), research logic (inductive and deductive), research purpose (explanatory, descriptive, exploratory, and evaluation), and research underpinning (positivist, interpretivist, constructivist, critical) as the strategy phase as input to choosing the concrete approaches for obtaining the results. These are strongly linked with the above questions. 
\\

\begin{mybox}
    \textbf{Teaching Module 3}
    
    \textbf{Intended Learning Outcome:} Develop the ability to formulate and critically evaluate research questions based on given research problem formulations and goals.

    \textbf{Teaching and Learning Activities:} Provide students with examples of software engineering problems and associated goals to perform a problem formulation exercise. Ask students to critique and discuss the clarity and focus of these formulations, encouraging collaboration and critical thinking.
    
    Then, have them develop a set of research questions that would effectively address the given problem and associated goal. As a role-playing activity, students take on the roles of different stakeholders (e.g., researchers and industry professionals) to formulate research questions from various perspectives.
    
    \textbf{Assessment:} Students submit the set of research questions they developed, assessing clarity, focus, and alignment with the initial problem.
\end{mybox}

\subsection{Frameworks for Formulating Research Questions}

While the above distinction of question types helps with methodological choices, the process of formulating well-defined RQs often requires additional support from specific frameworks. These frameworks are especially useful for novice researchers, as they help in thoroughly considering all aspects of the research. Key among these are the PICO framework and the Goal-Question-Metric (GQM) approach.

\textbf{PICO Framework:} Originally from healthcare research and often adopted in SLRs, the PICO framework aids in formulating precise research questions by focusing on four essential elements \cite{kitchenham2015evidence}: 

\begin{itemize}
    \item Population: Who are the subjects of the study?
    \item Intervention: What is the action or technology being investigated?
    \item Comparison: What is the intervention being compared to?
    \item Outcome: What are the expected effects or outcomes of the intervention? 
\end{itemize}

The PICO framework is particularly useful for empirical studies looking at the impact of specific technologies or methodologies. For instance, to investigate the impact of agile methodologies on project success rates, a research question might be, \textit{"In large tech companies (population), how does the adoption of Agile methodologies (intervention) compare to the traditional Waterfall model (comparison) in affecting project delivery times and productivity (outcomes)?"}

\textbf{GQM Approach:} Tailored for the SE domain by Basili et al. \cite{basili1994goal}, the GQM approach is particularly useful for evaluating technologies and strategic decisions in SE. It involves a three-step process: (1) define overall goals to be achieved; (2) establish specific questions that reflect the goal; and (3) determine the metrics needed and how data will be collected to answer the question. The stepwise approach ensures that all the elements are mapped, and a chain of evidence is established.

GQM is better suited for process improvement studies in SE. For example, if the goal is to assess the effectiveness of CI/CD practices, a suitable question might be, \textit{"What is the reduction in deployment failures observed after implementing CI/CD practices in software development teams?"} To address this question, one would measure key performance indicators (metrics) such as the deployment frequency, success/failure rates, and mean time to recovery, both before and after CI/CD implementation.

\textbf{Other Frameworks:} The FINER criteria evaluates the feasibility, interest, novelty, ethics, and relevance of research questions \cite{hulley2001conceiving}. The SPIDER Tool (standing for Sample, Phenomenon of Interest, Design, Evaluation, and Research type), offers a multifaceted approach to question formulation for mixed-methods research \cite{cooke2012beyond}.

Frameworks such as the ones listed above may not be ideally suited for interpretivist and constructivist research initially. These researches prioritizes understanding the meanings, experiences, and complexities of human interactions rather than measurement or prediction. In these paradigms, RQs are broad and open-ended at the initial stages to allow for an unrestricted exploration of the topic. The 5W1H method (Who, What, When, Where, Why, and How) supports the development of such questions and enhance their communication with a broader audience \cite{tattersall2015using}.

As the research progresses, these broad RQs evolve in response to new data and insights, capturing the dynamic nature of human behavior. For instance, a RQ might begin as \textit{"What processes influence software development practices in start-up environments?"} and evolve into \textit{"How do agile practices impact team dynamics in tech start-ups?"} Frameworks such as PICO, GQM, FINER and SPIDER become more relevant in the later stages of research, helping to refine and structure the developed theories into more precise research questions. 

\section{Framing a Research Approach}

Since our emphasis is on empirical research, we prioritize applied research over theoretical reasoning within this context. However, before connecting the research questions to the research logic, the purpose of the research and the research approach, we first briefly introduce the concepts. 

\begin{figure}[!ht]
\centering
    \includegraphics[width=\textwidth]{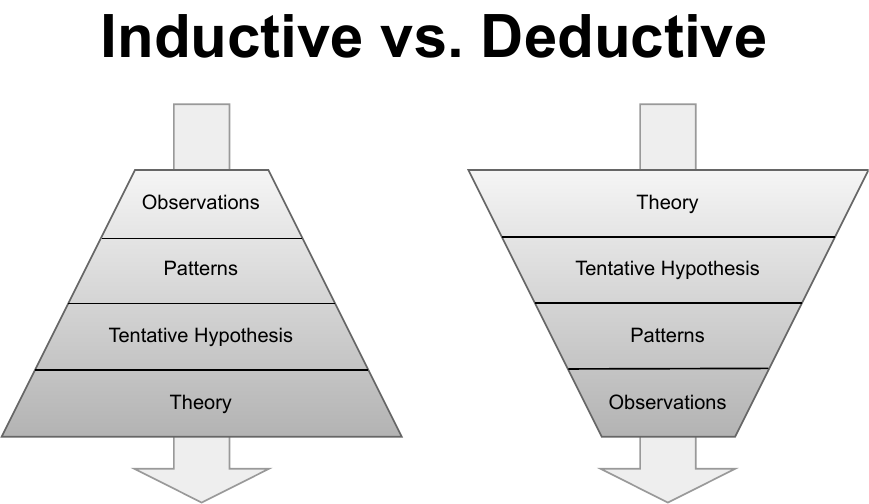}
    \caption{Inductive versus deductive research logic.}
    \label{fig:inductive-vs-deductive}
\end{figure}

\textbf{Research logic:} \textit{Inductive reasoning} involves making generalizations based on specific observations or experiences. This approach starts with specific data or instances and then draws broader generalizations or theories from them. \textit{Deductive reasoning}, on the other hand, works from the general to the specific. It starts with a general statement or hypothesis and conducts studies to test the hypothesis. An example is shown in Figure \ref{fig:inductive-vs-deductive}. \textit{Abductive reasoning}, is a third form of logical inference that seeks the simplest and most likely explanation from a set of observations. It useful for generating hypotheses that can be subsequently tested through inductive or deductive methods. Hypotheses may also be generated from descriptive and relationship questions (induction), and later be tested with causality questions (deduction).

\textbf{Research purpose:} \textit{Explanatory research} aims to clarify why and how there is a relationship between two aspects of a study. It focuses on understanding the causality of relationships and often follows up on findings from exploratory or descriptive research. \textit{Descriptive research} seeks to accurately and systematically describe a population, situation, or phenomenon. \textit{Exploratory research} is conducted for a problem that has not been clearly defined yet. That is, exploratory research aims to gain new insights and to formulate propositions and hypotheses, or more concrete research questions. \textit{Evaluation research} is concerned with, for example, assessing the effectiveness and efficiency of a software engineering intervention.

\textbf{Research underpinnings:} \textit{Positivism} assumes a single, objective reality that can be observed and measured through scientific methods. \textit{Interpretivism}, also known as constructivism, posits that our perception of reality and to a certain extend reality itself is shaped by human and social experience and interpretation. The \textit{Critical approach} views reality as shaped by, for example, social, cultural, or economic values with a focus on critically evaluating phenomena in that reality. \textit{Pragmatism} draws from the other underpinnings and often from abductive reasoning to emphasize that the validity or knowledge is determined by its practical utility in explaining or improving human practices.

The interconnection between research questions and the characteristics of research logic, purpose, and underpinnings is key for framing a research approach. We illustrate this interplay with examples from SE research, explaining how research logic, purpose, and underpinning align to address research questions:

\begin{enumerate}
    \item[RQ1:] \textit{What processes are emerging in DevOps teams to handle continuous integration and continuous deployment (CI/CD)?} \\ This open-ended \textbf{descriptive-process} question is suitable for an \textbf{exploratory}-purposed study that aims to generate insights via \textbf{inductive} reasoning . It aligns with an \textbf{interpretivist} underpinning as it seeks to understand emerging practices from the perspective of practitioners.
    \item[RQ2:] \textit{What are the most common types of bugs encountered in Agile software development projects?} \\ This \textbf{description and classification} question is grounded on a \textbf{positivist} underpinning. It uses \textbf{inductive} reasoning to draw on observed data, with a \textbf{descriptive} purpose of identifying the types of bugs based on empirical evidence.
    \item[RQ3:] \textit{Does implementing systematic code reviews reduce the frequency of bugs in software releases?} \\ This aims to \textbf{explain} the relationship between code reviews and bug frequency, making it a \textbf{causality} question. It uses \textbf{deductive} reasoning, starting with a hypothesis to be tested, and aligns with a \textbf{positivist} underpinning for objective measurement.
    \item[RQ4:] \textit{How effective is a given new project management tool in improving team productivity compared to the previous tool?} \\ This \textbf{causality-comparative} question uses \textbf{deductive} reasoning to \textbf{evaluate} the impact of the tool. The \textbf{critical approach} underpinning examines the effectiveness in an organizational context.
\end{enumerate}

It's important to note that research can employ a variety of combinations beyond those depicted in our examples. This inherent flexibility in research design allows researchers to effectively navigate the complexities of framing their specific research approach, facilitating the identification of suitable research question types (see Section \ref{sec:research-questions}). Vice versa, by formulating specific research questions, researchers can ensure that the research approach is consistent with the purpose, logic and underpinning. 
\\

\begin{mybox}
    \textbf{Teaching Module 4}
    
    \textbf{Intended Learning Outcome:} Develop the ability to characterize research regarding its logic, purpose, and underpinning.
    
    \textbf{Teaching and Learning Activities:} Assign students articles from ESE journals for research article analysis. Have them analyze and present the research logic (inductive or deductive), purpose (exploratory, explanatory, etc.), and underpinning (positivist, interpretivist, etc.) used in these articles.
    
    Also, in groups, students role-play as researchers planning a study. Each group is assigned a different combination of research logic, purpose, and underpinning, and they must outline a research design that aligns with these parameters.
    
    \textbf{Assessment:} Students’ article analyses and/or research plans are evaluated on the understanding and application of research logic, purpose, and underpinning
\end{mybox}

\section{Choosing Research Methods}

Next step into the research design is choosing the appropriate methodologies and techniques that will be used for collecting, sampling, and analyzing the data. There are multiple decision aspects in this process \cite{wohlin2015towards}, and they are highly influenced by the relevant empirical problem, knowledge gaps and research approach established earlier.

The research process itself can be viewed as a series of interconnected steps, each employing methods and techniques with differing scopes. These range from comprehensive processes that cover all aspects of data handling to more specific techniques focused solely on one aspect of the process \cite{stol2018abc}.

In this book, we have adopted a classification that aligns with the corpus of empirical studies in SE. It's important to recognize, however, that there are varied opinions among researchers regarding the classification of specific methods. In the literature, it was pointed out that methods are often misidentified by authors \cite{rainer2023case}, reflecting the complexity of method selection. See, for example, the works of Stol and Fitzgerald \cite{stol2018abc}, and Molléri, Petersen and Mendes \cite{molleri2019cerse}; both studies employ differing approaches in classifying research methods in ESE. The former adopts a taxonomy from more mature social sciences whereas the later uses an inductive approach based solely on how SE authors label their own studies

\subsection{Overview of Common SE Research Methods}
\label{sec:research-methods}

This section outlines common research methods in ESE, focusing on the specific contexts and scenarios where they are most effective, along with their requirements and practical considerations. It's designed to help researchers select the most suitable process for their research approach, fulfilling the methodological requirements and effectively linking them with specific research questions. Here, \textit{requirements} refer to a short checklist that helps determine whether a study indeed meets its methodological standards; more comprehensive checklists supporting peer-review can be found in Ralph et. al \cite{ralph2020empirical}. The research methods are organized according to the chapters of this book where they are further detailed.

\subsubsection{Experiment}

Experimental studies are pivotal in ESE research, particularly for evaluating the applicability of methods and technologies. These studies often involve controlled settings where participants (like practitioners, organizations, or even software systems) are exposed to different treatments to observe the effects on measurable variables. Randomized Controlled Trials (RCTs) are the gold standard for experimental studies, but other more reliant quasi-experiment designs are also common in SE research. Field experiments are conducted in real-world settings, which adds complexity and reduces control over variables compared to traditional experiments. Further information on experiments can be found in Chapter 14.

\textbf{Requirements:} The requirements on RCTs are relatively strict. The following questions help you evaluate whether your study is indeed a RCT.

\begin{todolist}
    \item Does the study involve the random assignment of participants or subjects to control and experimental groups?
    \item Is there a control group that does not receive the experimental treatment or intervention?
    \item Does the study investigate the effects of a specific intervention or treatment?
    \item Are procedures, treatments, and measurements standardized and applied consistently across all participants?
    \item Are specific outcomes being measured to assess the effects of the intervention?
    \item Are participants, researchers, or both blinded to group assignments (single-blind or double-blind)?
    \item Is statistical analysis used to compare outcomes between control and experimental groups?
\end{todolist}

While quasi-experiments and field studies can be assessed using a similar checklist, they differ from RCTs in that random assignment and strict control over procedures (i.e., the first two questions) may not be feasible. In addition, blinding of participants and researchers (i.e, question 6) is less strict in quasi-experiments and not practical in real-world settings.

\subsubsection{Simulation}

Simulation offers a valuable approach for exploring complex systems, processes, or behaviors in a controlled, virtual environment. This method is used when real-world experimentation is impractical or impossible. Simulations allow researchers to model and analyze software development scenarios, including system performance, user behavior, or project management strategies. They provide insights through the creation and testing of virtual models, which can replicate various aspects of SE processes. Detailed information and guidelines to simulation-based studies are provided in Chapter 16.

\textbf{Requirements:} The following questions help deciding whether it is a simulation study: 

\begin{todolist}
    \item Is a realistic and representative model of a process or system developed for the simulation?
    \item Are the parameters and variables clearly defined and based on empirical data or credible assumptions?
    \item Does the simulation allow for controlled manipulation and observation of key variables?
    \item Are the expected outcomes of the simulation aimed at drawing conclusions about real-world practices or theories?
    \item Is there a process for validating the simulation results against actual scenarios or data?
\end{todolist}

\subsubsection{Action research}

Action research is a method in which researchers actively engage in real-world problem-solving and practice improvement while acquiring knowledge. This approach, characterized by iterative planning, action, observation, and reflection, is particularly suitable for exploring transformations in software processes, such as adopting new technologies or modifying existing processes or redefining organizational structures.  ey to action research is the active participation of both researchers and software professionals, bridging theory and practice. Best employed for testing and refining methods and tools in live settings, action research is further detailed in Chapters 17 and 18.

\textbf{Requirements:} The following questions help deciding whether the planned study is indeed an action research: 

\begin{todolist}
    \item Does the study aims at solving a specific practical problem within a real-world setting?
    \item Does the research involve collaboration between researchers and participants who are directly affected by or interested in the problem?
    \item Does the study use an iterative process, involving cycles of planning, acting, observing, and reflecting?
    \item Does the research combine practical action with generating theoretical knowledge about the problem and its solutions?
    \item Is the research design flexible and adaptive to changes and new findings as the study progresses?
\end{todolist}

\subsubsection{Case Study}

Case study research delves into a specific research problem within their real-life context, often focusing on organizations, projects, or processes. This method allows handling complex phenomena through triangulation, which involves integrating various data sources for deeper insights. Unlike action research or design science, the researcher in a case study observes without direct involvement in the process. Designing a case study requires careful consideration of context, as highlighted by Petersen and Wohlin's \cite{petersen2009context} framework. Particularly effective in situations where the  boundaries between the phenomenon and context are blurred, case studies are discussed more extensively in Chapter 19.

\textbf{Requirements:} The following questions help in determining whether a study is a case study, as discussed in Wohlin and Rainer \cite{wohlin2021case}:

\begin{todolist}
    \item Does the study describe a case? Examples for cases could be a software system, a development process, a software organization, etc.
    \item Does the study investigate a contemporary phenomena (occurs at present time)?
    \item Is the study conducted in a real life context (e.g., industry, open source), thus involving real developers, products and processes?
    \item Are multiple methods for data collection used (triangulation)?
    \item Does the researcher act passively (only collecting data) and therefore not take an active role in the investigated case?
\end{todolist}

\subsubsection{Design Science}

Design Science is ideal for creating innovative solutions to specific problems where existing theories are insufficient. It's about actively developing, testing and refining new tools, methods, or models in real-world environments. Similar to action research, this method encourages a problem-solving approach and collaboration with practice, but it focuses on creating a practical and tangible solution. Its outcomes contribute significantly to advancing SE practices and methods, bridging a gap between theory and practice. Further details are provided in Chapter 20.

\textbf{Requirements:} For assessing if a study is Design Science, use the following questions: 

\begin{todolist}
    \item Does the study aim to create a novel artifact (tool, method, model) to solve a specific problem?
    \item Is the artifact designed based on identified requirements and knowledge gaps?
    \item Does the research involve iterative cycles of development and evaluation of the artifact?
    \item Are the results of artifact evaluation used to refine and improve its design?
    \item Does the study contribute both to practical applications and theoretical understanding?
\end{todolist}

\subsubsection{Survey}

Survey is a research method for gathering data on practices, opinions, or behaviors from a broad audience, often practitioners. They are particularly effective in understanding trends, attitudes, and patterns within the community. This method is ideal to collect extensive data from a large population, making it particularly useful for understanding the prevailing practices in SE. It can also effectively compare different subgroups within the population. Surveys can be tailored to specific research questions, making them versatile for various knowledge areas. Further details about surveys are given in Chapter 21.

\textbf{Requirements:} The following questions help determine if a study is a survey:  

\begin{todolist}
    \item Is the survey designed to collect data on practices, opinions, or behaviors from a wide audience?
    \item Are the survey questions aligned with the research objectives and hypotheses?
    \item Is the survey sample representative of the target population?
    \item Is there a clear plan for data collection and analysis of survey results appropriate for the study design?
    \item Does the study ensure ethical considerations, such as consent and anonymity of respondents?
\end{todolist}

\subsubsection{Literature Review}

Systematic Literature Review (SLR) is a comprehensive overview of existing literature on a specific topic, ensuring a thorough understanding of the current state of the art. SLRs follow a structured approach to minimizing bias in literature selection and analysis. They are crucial for identifying research gaps, summarizing existing evidence, and setting the foundation for future research. Systematic Mapping Studies (SMS) are broader in scope but less focused than SLRs, providing a broader overview of a research field and categorizing existing literature. Multivocal Literature Review (MLR) includes both academic and grey literature, offering a perspective from both research and practice. These methods are invaluable for consolidating knowledge in SE. More on this research method is detailed in Chapters 22 and 23.

\textbf{Requirements:} The following questions help deciding whether the planned study is indeed a SLR:  

\begin{todolist}
    \item Is the review guided by a clear set of research questions or objectives?
    \item Does it follow a predefined and transparent methodology for searching, selecting, and appraising literature?
    \item Are inclusion and exclusion criteria for literature selection clearly defined and applied consistently?
    \item Are the credibility and relevance of literature sources critically assessed?
    \item Does the review synthesize findings from selected literature to answer the research questions?
    \item Is the review's methodology documented in a manner that allows for future replication and validation?
\end{todolist}

SMS can be assessed using a similar checklist, but the synthesis approach (see question 4) in SMS is about mapping the field's landscape rather than providing detailed answers to research questions. For MLR, we suggest adding the additional requirements:

\begin{todolist}
    \item Does the review include a mix of academic and grey literature (such as industry reports, blogs, and whitepapers)?
    \item Is there a strategy to integrate and compare insights from both types of sources?
\end{todolist}

\subsubsection{Ethnography}

In ESE, ethnography studies focus on understanding the cultural and social dynamics within live environments, such as a software project or a development team. It is a qualitative research method involving immersive observation and interaction with participants in their natural settings, gathering insights into the behaviors, practices, and interactions. It aims at uncovering implicit knowledge, workflows, and challenges that might not be evident through other research methods. Ethnography is useful when exploring the human and organizational aspects of software development, offering rich, detailed insights into the lived experiences of practitioners. Chapter 24 offers more details on ethnographic studies.

\textbf{Requirements:} The following questions help in designing an ethnography:

\begin{todolist}
    \item Does the study involve immersive, long-term observation within a realistic setting?
    \item Are qualitative data collection methods like participant observation and interviews used?
    \item Does the researcher engage deeply with the cultural and social context of the studied environment?
    \item Is there a focus on investigating human or organizational factors, such as behaviors, practices, and interactions of participants?
\end{todolist}

\subsubsection{Grounded Theory}

Grounded Theory (GT) is a qualitative research method aimed at generating theories from empirical data. Primarily used in exploratory studies, it involves systematic data collection and analysis to construct theoretical frameworks that explain the observed phenomena. GT involves systematic data collection and analysis, often through techniques such as interviews, observations, and document analysis. It is particularly useful for understanding complex human and organizational dynamics in software development, offering insights into processes, behaviors, and interactions within this field. 

\textbf{Requirements:} For assessing if a study is grounded theory, use the following questions: 

\begin{todolist}
    \item Does the study aim to develop a theory grounded in empirical data?
    \item Is there an iterative process of data collection and analysis?
    \item Are concepts and categories developed inductively from the data?
    \item Is there a constant comparison of data to refine and develop theoretical insights?
    \item Does the study emphasize theoretical sampling to explore emerging themes and concepts?
\end{todolist}

\begin{mybox}
    \textbf{Teaching Module 5}
    
    \textbf{Intended Learning Outcome:} Develop the ability to assess whether studies meet specified methodological requirements.

    \textbf{Teaching and Learning Activities:} Provide students with a set of criteria for different research methods. Assign a selection of research papers labeled as a specific type of study, and ask them to analyze each paper to determine if it meets the methodological requirements of the given study type.

    Also, students role-play as authors and peer-reviewers in a research method debate. The 'authors' defend their methodology choice, while 'peer reviewers' critically assess the alignment with method requirements.

    \textbf{Assessment:} Students’ analyzes are evaluated on the accurate assessment of  methodological adherence and argumentation on methodological validity.
\end{mybox}

\begin{mybox}
    \textbf{Teaching Module 6}

    \textbf{Intended Learning Outcome:} Develop the ability to argue for and against the choice of research methods based on a formulated research strategy.
    
    \textbf{Teaching and Learning Activities:} Students are divided into groups, each advocating for a different research method as the best fit for a given research approach (including problem formulation, study goals and questions/hypotheses). A debate is held to discuss the merits and drawbacks of each method in relation to the strategy.
    
    \textbf{Assessment:} Students are assessed based on their participation in debates and their arguments in support of or against the methods discussed.
\end{mybox}

\subsection{Data Collection}
\label{sec:data-collection}

This section serves as a toolkit to various data collection techniques, outlining the applications, advantages, and the types of data that each technique yields. These methods address different research challenges in ESE, offering both breadth and depth in data collection strategies.

\subsubsection{Document Analysis}

This method involves examining and interpreting documents to gather data. It is essential in SLRs, case studies, and sometimes in action research. In literature reviews, researchers systematically search and select academic papers and other scholarly articles to understand the existing body of knowledge on a topic. In field studies, researchers gather a variety of documents such as technical specifications, project reports, user manuals, and development logs. This method offers insights into the historical and contextual background of a research topic, contributing to a more comprehensive understanding of it.

\subsubsection{Data Mining}

Data mining or Mining Software Repositories (MSR) involves analyzing extensive data sets to uncover patterns, trends, and anomalies. It's particularly aligned with experimental studies and design science research. Data mining can reveal insights into development practices, code quality, and team productivity. By applying statistical and machine learning techniques, researchers can extract meaningful information from large volumes of data, offering a data-driven understanding of processes and outcomes. Further details on MSR can be found in Chapter 15.

\subsubsection{Observations}

Observations as a data collection method involve directly watching processes, practices, and behaviors as they naturally occur in real-world settings. This method is useful in case studies, action research, and ethnography, where understanding the workflow and interactions in software development environments is vital. Through observations, researchers capture nuances and dynamics that might not be revealed through direct inquiry. It allows for a deeper understanding of phenomena such as team collaboration, decision-making processes, and the actual use of tools and methods in practice, providing rich, contextual insights into the everyday realities.

\subsubsection{Questionnaires}

Questionnaires are a tool for systematically collecting data from a large number of respondents. It is particularly effective in survey research, where understanding broad trends, opinions, or behaviors across a wide audience is essential. Questionnaires are used to gather quantitative and qualitative data on various aspects, such as developer experiences, user satisfaction, or tool effectiveness. Their structured format allows for efficient data analysis, and they can be distributed widely, making them ideal for large-scale studies. The design of questionnaires, including question clarity and response format, is crucial for ensuring accurate and useful responses.

\subsubsection{Interviews}

Through direct conversation with individuals, researchers can gather detailed information about their experiences, opinions, and practices. Interviews are especially beneficial in case studies and action research, where understanding personal perspectives is crucial. It allows researchers to explore complex topics like decision-making processes, experiences with specific technologies, or perceptions of industry trends. This method facilitates a deeper, more nuanced understanding of the human elements, often uncovering information that isn't accessible through quantitative methods. With regard to interviews, the type of interview to be conducted needs to be decided:

\begin{itemize}
    \item \textit{Structured interviews} with a standardized set of questions that are strictly followed. 
    \item \textit{Unstructured interviews} which are informal and do not follow a set of predetermined questions.
    \item \textit{Semi-structured interviews} mixing predetermined questions and those emerging during the conversation.
\end{itemize}
 
\subsubsection{Other data collection methods}

Beyond the commonly mentioned methods, there are other data collection techniques worth noting. \textbf{Focus groups and workshops} involve gathering a group of individuals to discuss and provide feedback on a particular topic or issue. \textbf{Delphi studies} consist of rounds of inquiry to a panel of experts, with each round using feedback to refine opinions and reach consensus on specific topics. \textbf{Artifact analysis} involves examining the software products, such as code quality, architecture, and usability. These methods provide additional lenses through which researchers can gather valuable data, complementing more traditional data collection approaches.
\\

\begin{mybox}
    \textbf{Teaching Module 7}
    
    \textbf{Intended Learning Outcome:} Develop the ability to evaluate and select appropriate data collection strategies for various SE research contexts.

    \textbf{Teaching and Learning Activities:} Organize a data collection workshop where students must analyze different data collection methods, discussing their advantages, limitations, and applicability to various SE research contexts. 

    In addition, assign students to select a data collection strategy for a given research approach, requiring them to justify their choice based on the research objectives and context.

    \textbf{Assessment:} Students’ justifications are evaluated on the understanding and application of data collection methods.
\end{mybox}

\subsection{Data analysis}
\label{sec:data-analysis}

This section overviews various data analysis methods used in ESE research, including both qualitative and quantitative approaches. This summary helps in understanding the strengths of each approach and how they can be applied to analyze data effectively.

\subsubsection{Thematic Analysis}

Thematic analysis is another qualitative qualitative approach, used for identifying and interpreting patterns or themes in data. It's particularly useful in organizing and analyzing qualitative data like interview transcripts and observation notes. This method helps in understanding the underlying processes and dynamics of the phenomenon under study. ESE often employs thematic analysis to gain insights into areas like developer experiences, team interactions, or user feedback.

\subsubsection{Content Analysis}
Content Analysis, offers a more structured approach to analyzing textual data. It involves quantifying and analyzing the presence, meanings, and relationships of certain words, themes, or concepts. Researchers use content analysis to systematically evaluate communication patterns, document contents, or online discussions. This method is particularly valuable for assessing trends, attitudes, and cultural norms within the SE community, providing a quantitative perspective on qualitative data. 

\subsubsection{Statistical analysis}

This is a fundamental tool for interpreting quantitative data and turning it into actionable insights. The analysis involves using statistical techniques to test hypotheses, understand relationships, and draw conclusions from data. This can include descriptive statistics to summarize data, inferential statistics to make predictions or inferences from a sample to a population, correlational analysis to assess relationships between variables, and advanced techniques like regression analysis to understand the relationship between variables. Used in experimental studies, surveys, and sometimes in case studies.

\subsubsection{Machine Learning}

Machine learning (ML) as a data analysis approach involves using algorithms to analyze large datasets and extract insights. It's effective in identifying patterns, predicting outcomes, and automating decision-making processes. ML can be applied to areas such as defect prediction, code analysis, and user behavior modeling. This approach transforms large, complex datasets into meaningful information, aiding in the development of more efficient and effective software engineering practices.

\subsubsection{Meta analysis}

Meta-analysis is a quantitative method used to statistically combine and analyze results from multiple studies on a similar topic. It provides a comprehensive view of the existing evidence, identifying overall trends and drawing more robust conclusions than individual studies alone. Meta-analysis is particularly valuable for synthesizing findings across different research projects, thereby enhancing the understanding of the effectiveness and impact of various SE practices and tools. This method offers a systematic approach to assess the generalizability of findings and to identify areas where research is consistent or where there are discrepancies.

\subsubsection{Other data analysis methods}

Beyond the most frequently mentioned analysis techniques, there are others worth mentioning. \textbf{Cross-Case Analysis} is employed in case study research to compare findings across multiple cases, enhancing the generalizability by drawing comparisons and contrasts across several scenarios. \textbf{Iterative testing and Prototype analysis} involve evaluating and refining a product or process through repeated cycles, allowing for gradual improvements in proposed solutions. 
\\

\begin{mybox}
    \textbf{Teaching Module 8}
    
    \textbf{Intended Learning Outcome:} Develop the ability to select appropriate data analysis methods aligned with the data type and research needs.
    
    \textbf{Teaching and Learning Activities:}  In a workshop, students analyze various data types (e.g., survey results, interview transcripts) to choose the most suitable analysis method and justify their selections. They also compare methods which are similar or target the same type of data (e.g., content vs. thematic analysis) and discuss their suitability in specific research contexts.
    
    \textbf{Assessment:} Students are assessed on the rationale behind their choices, with emphasis on their ability to align data analysis methods with data types.
\end{mybox}

\subsection{Selecting Research, Data Collection and Analysis Methods}

Having briefly presented the different research methods, for the given questions, we have a first pointer which methods may be suitable for the questions asked. Table \ref{tab:mapping-question-types} illustrates how different research question types (see Section \ref{sec:research-questions}) can be matched with potential research methods, and provides a brief rationale for each method’s selection. Please note that this table is not comprehensive; it serves as an example, and many other valid method-question pairings exist.

\begin{table}[!ht]
\caption{Mapping of question types to research methods with rationale.}
\label{tab:mapping-question-types}
\rowcolors{1}{white}{lightgray}
\begin{tabular}{p{1.8cm}>{\raggedright}p{2.6cm}p{7cm}}
\hline\noalign{\smallskip}
\textbf{Question Type} & \textbf{Potential Research Methods} & \textbf{Argumentation for Usage} \\
\noalign{\smallskip}\svhline\noalign{\smallskip}
Existence &
Surveys; Observations &
Surveys are effective for quantifying prevalence, and observations allow for firsthand witnessing of the phenomenon in its natural context. \\
Description and Classification &
Document Analysis; Case Studies &
Document analysis is suitable for understanding documented properties and categorizations, while case studies provide an in-depth look at specific instances to understand their nature and classify them. \\
Descriptive-Comparative & 
Surveys; Case Studies &
Surveys can gather comparative data across a large sample, and case studies allow for a detailed comparison of specific instances. \\
Descriptive-Process &
Observations; Case Studies &
Observations give a real-time view of processes as they occur, while case studies can provide a comprehensive understanding of the process within a particular context. \\
Relationship &
Surveys; Interviews &
Surveys can quantify and identify relationships across a broad sample, and interviews allow for in-depth exploration of perceptions and experiences regarding these relationships. \\
Causality &
Controlled Experiments; Data Mining &
Controlled experiments are ideal for testing causal relationships under controlled conditions, while data mining can uncover patterns and potential causal links in large datasets. \\
Causality-Comparative &
Controlled Experiments; Surveys &
Controlled experiments accurately compare different conditions for causal effect, while surveys can compare perceptions and experiences across different groups to infer comparative causality. \\
Causality-Comparative Interaction &
Controlled Experiments; Data Mining &
Controlled experiments can analyze variable interactions under specific conditions, and data mining helps in identifying complex patterns of interaction in large-scale data. \\
Design &
Design Science; Action Research &
Design Science is suitable for developing and evaluating artifacts to meet desired goals, while Action Research allows for iterative testing and refinement in real-world settings. \\
\noalign{\smallskip}\hline\noalign{\smallskip}
\end{tabular}
\end{table}

To illustrate an example where a mismatch between research question types and research methods can lead to ineffective combinations, consider using surveys for answering causality questions. While effective for gathering a broad range of data, survey has limited control of confounding factors, which makes it unsuitable for establishing causal relationships. Still, surveys can be quite useful for Causality-Comparative Questions, where comparing perceptions or experiences across different groups can infer comparative causality.

The second step in research involves aligning research methods with suitable data collection and analysis techniques. This strategic pairing is essential for ensuring the effectiveness and validity of the research. As Wohlin and Aurum \cite{wohlin2015towards} suggest, research methods should be viewed as frameworks incorporating various data handling techniques. Table \ref{tab:methods-pairing} provides typical pairings for ESE, but researchers are encouraged to explore alternative combinations tailored to their specific research needs and contexts. This adaptable approach ensures a more targeted and effective research methodology.

\begin{table}[!ht]
\caption{Typical combinations of research methods and data collection and data analysis approaches. The data techniques are ordered by their significance to the research method.}
\label{tab:methods-pairing}
\rowcolors{1}{white}{lightgray}
\begin{tabular}{p{3cm}>{\raggedright}p{4.4cm}p{4cm}}
\hline\noalign{\smallskip}
\textbf{Research Method} & \textbf{Data Collection} & \textbf{Data Analysis} \\
\noalign{\smallskip}\svhline\noalign{\smallskip}
Action Research &
Observations; Interviews; Document analysis; Focus groups &
Thematic analysis; Statistical analysis (descriptive, inferential) \\
Survey &
Questionnaires; Interviews &
Statistical analysis (descriptive, inferential, correlational) \\
Literature Reviews &
Document analysis &
Content analysis; Meta-analysis \\
Experiment &
Data mining; Questionnaires; Artifact analysis &
Statistical analysis (inferential, correlational, regression) \\
Simulation &
Data mining; Observation; Artifact analysis &
Statistical analysis (descriptive, inferential) \\
Ethnography &
Observation; Interviews; Artifact analysis &
Thematic analysis \\
Case Study &
Interviews; Document analysis; Observations, Questionnaires; Focus groups &
Thematic analysis; Content analysis; Cross-case analysis; Statistical analysis (descriptive) \\
Design Science &
Data mining; Observations; Document analysis; Delphi study; Artifact analysis &
Iterative testing; Prototype analysis \\
Grounded theory &
Observation; Interviews; Document analysis &
Thematic analysis; Content analysis \\
\noalign{\smallskip}\hline\noalign{\smallskip}
\end{tabular}
\end{table}


Moreover, the application of triangulation in research, which involves using multiple data sources, methods, or theoretical perspectives, further enhances the credibility and depth of findings. In ESE, triangulation allows researchers to validate results through different lenses, cross-verify findings, identify inconsistencies, and capture various dimensions of the research problem. For instance, when studying software development practices, combining quantitative surveys to gather broad data on team productivity with qualitative interviews to gain in-depth insights into individual experiences can offer a richer and more nuanced understanding. This approach ensures a more robust analysis that better reflects the complexities of real-world contexts.

\begin{mybox}
    \textbf{Teaching Module 9}
    
    \textbf{Intended Learning Outcome:} Develop the ability to synergize data collection and analysis methods to align with research questions, methods, and data types.
    
    \textbf{Teaching and Learning Activities:} Students role-play as researchers where they must choose an overall research method, and suitable data collection and analysis techniques for a given research problem, explaining how these choices complement each other. Utilize collaborative tools like Miro or Trello to allow students to visually map out research designs.
    
    
    \textbf{Assessment:} Students receive real-time feedback from peers and instructors on their collaborative tools. They are assessed on how well the methodological choices align with the research questions, methods and data types.
\end{mybox}

\subsection{Data Collection and Analysis in Flexible Research Designs}

Flexible research designs, characterized by their iterative nature, present unique challenges and opportunities in the context of data collection and analysis. Data collection methods are not static but evolve based on insights gained during the research process. Initial methods are chosen to address preliminary questions, but as new patterns or themes emerge from the data, researchers may expand their data collection techniques. This iterative approach includes emergent sampling, i.e. including additional participants or sources as needed; and triangulation, i.e. employing multiple data collection methods to uncover new dimensions of the research problem.

Data analysis in flexible designs is an ongoing process that begins early and evolves throughout the study. Researchers continuously reflect on data, refining research questions and adjusting methods as new insights emerge. This iterative analysis, knows as progressive focusing, involves broad initial data examination that gradually narrows to focus on significant emerging themes. Effective implementation of flexible data collection and analysis allows researchers to align their analytical focus with the evolving research context and findings.

\section{Ensuring Feasibility and Relevance}

While the questions we wish to answer should be the main driver for the method selected, it may not always be feasible. This section explores additional factors influencing method selection, such as practical constraints and the study's relevance to current SE problems. Emphasis is given to trade-offs inherent in these choices and how they impact the decision-making process, ensuring researchers can balance theoretical aims with practical realities effectively.

Here, we would like to encourage researchers and students to be aware of such factors and make them explicit when arguing and discussing the choice of research methods, as there may be practical reasons for adjusting the choice of methods besides the research questions. They, however, should be the main driver and implications of deviations need to be well founded.

\subsection{Methodological implications}

Research methods vary in their focus. Surveys, for example, provide breadth, capturing wide-ranging data from large samples, ideal for general trends. In contrast, case studies offer depth, providing detailed insights into specific scenarios or organizations. Researchers must decide between a broad overview of a topic and an in-depth understanding of a particular aspect.

Research methods also differ in how obtrusive they are to participants and their applicability in specific contexts. For instance, ethnographic studies are less obtrusive, allowing observation of natural behaviors, but are highly context-specific. Surveys, while more obtrusive, are less context-dependent and can be applied more universally \cite{stol2018abc}.

Context includes environmental factors, organizational settings, and specific conditions under which the software engineering practices are carried out. Understanding the context helps in designing studies that are not only theoretically sound but also practically applicable. This involves considering specific nuances of the software development environment and tailoring research methods to suit these unique characteristics. Contextual awareness ensures that findings are reflective of, and applicable to, the real-world scenarios they intend to address.

Finally, relevance is a critical factor in ESE research. It pertains to the significance and applicability of the research to real-world problems. A relevant study addresses current challenges or gaps in SE practice, thereby contributing meaningful insights. Researchers must continually assess the relevance of their work, ensuring it aligns with current industry trends, technological advancements, and practical needs.

\subsection{Practical implications}

Wohlin and Aurum \cite{wohlin2015towards} pointed out that factors beyond methodological ones may also influence the method choice. As an example, they present a case where a student chooses a method based on the competence of using the method. Effective research design requires acknowledging practical limitations, such as time and resources. Researchers are encouraged to realistically evaluate their capabilities against the demands of the chosen methods.

Data availability also plays a crucial role; for instance, case studies demand detailed data from specific organizations, while extensive surveys need access to large audiences. Obviously, conducting a case study requires good connections to industry. Also, we may not always be able to obtain the sample in surveys we ideally would like to have. Practically, random sampling would be desirable, but in software engineering the most common approaches for sampling are purposive and convenience sampling \cite{baltes2022sampling}. 

Choosing a method that demands extensive time and resources may not be feasible for a small team with limited funding. For instance, longitudinal studies evaluating the long-term effects of implementing new practices, require significant investment in time and personnel. Similarly, opting for depth, such as conducting detailed ethnographic studies to understand developer interactions within teams, may limit the generalizability due to their focus on specific environments.

The trade-offs between these elements significantly impact the research design decision-making process. Flexible designs, in particular, allow for adjustments in response to changing circumstances, making the research more responsive to evolving data and insights. This flexibility is key for research topics where the scope is likely to shift, ensuring that studies reflect current trends. Researchers must carefully weigh these trade-offs to ensure their chosen methodology aligns with their research goals and practical constraints.

\section{Summary}

This chapter is a foundational guide for SE researchers on designing research proposals, setting the stage for further chapters in this book. It focuses on identifying relevant empirical problems, formulating targeted research questions, and choosing appropriate methodologies. It addresses challenges in method selection and classification within ESE and provides insights into various methods for data collection, and analysis techniques. The chapter aims to help researchers make informed methodology choices and understand their research's practical implications. It also includes interactive and engaging learning activities for a high-education setting. The following chapters will expand onto research methods and strategies, provide in-depth insights and hands-on guidance for teaching empirical research methods in SE.

\subsection{Further Reading}

Other chapters of the book "Teaching Empirical Research Methods in Software Engineering" mentioned in this document are:

\begin{enumerate}
    \item[Chapter 14] \textbf{A Course on Experimentation in Software Engineering: Focusing on Doing}, by Sira Vegas, Natalia Juristo
    \item[Chapter 15] \textbf{Teaching Mining Software Repositories}, by Zadia Codabux. Fatemeh Fard, Roberto Verdecchia, Fabio Palomba, Dario Di Nucci, Gilberto Recupito 
    \item[Chapter 16] \textbf{Teaching Simulation as a Research Method in Empirical Software Engineering}, by Breno Bernard Nicolau de França, Nauman bin Ali, Dietmar Pfahl, Valdemar Vicente, Graciano Neto 
    \item[Chapter 17] \textbf{Teaching Action Research} by Mirosiaw Staron 
    \item[Chapter 18] \textbf{Action Research with Industrial Software Engineering - An Educational Perspective} by Yvonne Dittrich, Johan Bolmsten, Cathrine Seidelin 
    \item[Chapter 19] \textbf{Teaching Case Study Research} by Stefan Wagner 
    \item[Chapter 20] \textbf{Teaching Design Science as a Method for Effective Research Development} by Oscar Pastor, Mmatshuene Anna Segooa, Jose Ignacio Panach 
    \item[Chapter 21] \textbf{Teaching Survey Research in Software Engineering} by Marcos Kalinowski, Allysson Allex Araújo, Daniel Mendez 
    \item[Chapter 22] \textbf{Teaching Literature Reviews in Software Engineering Research} by Sebastian Baltes, Paul Ralph 
    \item[Chapter 23] \textbf{Teaching Systematic Literature Reviews: Strategies and Best Practices} by Marcela Genero, Mario Piattini 
    \item[Chapter 24] \textbf{Teaching and Learning Ethnography for Software Engineering Contexts} by Yvonne Dittrich, Helen Sharp, Cleidson R. B. de Souza 
\end{enumerate}

In addition to this book's complementary chapters, readers are encouraged to explore Eastbrook et al. \cite{easterbrook2008selecting} for selecting empirical methods in SE research. Wohlin and Aurum's \cite{wohlin2015towards} decision-making framework, extensively referenced in this chapter, is a valuable resource we built upon. For comprehensive guidelines on various methods, Molléri, Petersen, and Mendes \cite{molleri2019cerse} offer an extensive compilation of methodological literature. Creswell's book \cite{creswell2017research}, while more general, provides insightful guidance for research design across various fields, aiding in formulating research proposals.

\bibliographystyle{abbrv}
\bibliography{references}

\begin{thebibliography}{10}

\bibitem{baltes2022sampling}
S.~Baltes and P.~Ralph.
\newblock Sampling in software engineering research: A critical review and guidelines.
\newblock {\em Empirical Software Engineering}, 27(4):94, 2022.

\bibitem{basili1994goal}
V.~R. Basili, G.~Caldiera, and H.~D. Rombach.
\newblock The goal question metric approach.
\newblock {\em Encyclopedia of software engineering}, pages 528--532, 1994.

\bibitem{benbasat1999empirical}
I.~Benbasat and R.~W. Zmud.
\newblock Empirical research in information systems: The practice of relevance.
\newblock {\em MIS quarterly}, pages 3--16, 1999.

\bibitem{biggs2022teaching}
J.~Biggs, C.~Tang, and G.~Kennedy.
\newblock {\em Teaching for quality learning at university 5e}.
\newblock McGraw-hill education (UK), 2022.

\bibitem{bourque2014swebok}
P.~Bourque and R.~E. Fairley.
\newblock Swebok v3. 0: Guide to the software engineering body of knowledge.
\newblock {\em IEEE Computer Society}, pages 1--335, 2014.

\bibitem{cooke2012beyond}
A.~Cooke, D.~Smith, and A.~Booth.
\newblock Beyond pico: the spider tool for qualitative evidence synthesis.
\newblock {\em Qualitative health research}, 22(10):1435--1443, 2012.

\bibitem{creswell2017research}
J.~W. Creswell and J.~D. Creswell.
\newblock {\em Research design: Qualitative, quantitative, and mixed methods approaches}.
\newblock Sage publications, 2017.

\bibitem{easterbrook2008selecting}
S.~Easterbrook, J.~Singer, M.-A. Storey, and D.~Damian.
\newblock Selecting empirical methods for software engineering research.
\newblock {\em Guide to advanced empirical software engineering}, pages 285--311, 2008.

\bibitem{garousi2019guidelines}
V.~Garousi, M.~Felderer, and M.~V. M{\"a}ntyl{\"a}.
\newblock Guidelines for including grey literature and conducting multivocal literature reviews in software engineering.
\newblock {\em Information and software technology}, 106:101--121, 2019.

\bibitem{hulley2001conceiving}
S.~B. Hulley, S.~R. Cummings, W.~S. Browner, D.~G. Grady, N.~Hearst, and T.~Newman.
\newblock Conceiving the research question.
\newblock {\em Designing clinical research}, 335, 2001.

\bibitem{kampenes2008flexibility}
V.~B. Kampenes, B.~Anda, and T.~Dyb{\aa}.
\newblock Flexibility in research designs in empirical software engineering.
\newblock In {\em 12th International Conference on Evaluation and Assessment in Software Engineering (EASE)}. BCS Learning \& Development, 2008.

\bibitem{kitchenham2015evidence}
B.~A. Kitchenham, D.~Budgen, and P.~Brereton.
\newblock {\em Evidence-based software engineering and systematic reviews}, volume~4.
\newblock CRC press, 2015.

\bibitem{molleri2019cerse}
J.~S. Moll{\'e}ri, K.~Petersen, and E.~Mendes.
\newblock Cerse-catalog for empirical research in software engineering: A systematic mapping study.
\newblock {\em Information and Software Technology}, 105:117--149, 2019.

\bibitem{petersen2024revisiting}
K.~Petersen, J.~B{\"o}rstler, N.~bin Ali, and E.~Engstr{\"o}m.
\newblock Revisiting the construct and assessment of industrial relevance in software engineering research.
\newblock In {\em 1st International Workshop on Methodological Issues with Empirical Studies in Software Engineering (WSESE'2024)}, 2024.
\newblock accepted.

\bibitem{petersen2009context}
K.~Petersen and C.~Wohlin.
\newblock Context in industrial software engineering research.
\newblock In {\em 2009 3rd international symposium on empirical software engineering and measurement}, pages 401--404. IEEE, 2009.

\bibitem{rainer2023case}
A.~Rainer and C.~Wohlin.
\newblock Case study identification: A trivial indicator outperforms human classifiers.
\newblock {\em Information and Software Technology}, page 107252, 2023.

\bibitem{ralph2020empirical}
P.~Ralph, N.~b. Ali, S.~Baltes, D.~Bianculli, J.~Diaz, Y.~Dittrich, N.~Ernst, M.~Felderer, R.~Feldt, A.~Filieri, et~al.
\newblock Empirical standards for software engineering research.
\newblock {\em arXiv preprint arXiv:2010.03525}, 2020.

\bibitem{robson2016real}
C.~Robson and K.~McCartan.
\newblock {\em Real World Research}.
\newblock Wiley, 2016.

\bibitem{stol2018abc}
K.-J. Stol and B.~Fitzgerald.
\newblock The abc of software engineering research.
\newblock {\em ACM Transactions on Software Engineering and Methodology (TOSEM)}, 27(3):1--51, 2018.

\bibitem{tattersall2015using}
A.~Tattersall.
\newblock Who, what, where, when, why: Using the 5 ws to communicate your research.
\newblock {\em Impact of Social Sciences Blog}, 2015.

\bibitem{wohlin2021case}
C.~Wohlin.
\newblock Case study research in software engineering—it is a case, and it is a study, but is it a case study?
\newblock {\em Information and Software Technology}, 133:106514, 2021.

\bibitem{wohlin2015towards}
C.~Wohlin and A.~Aurum.
\newblock Towards a decision-making structure for selecting a research design in empirical software engineering.
\newblock {\em Empirical Software Engineering}, 20:1427--1455, 2015.

\end{thebibliography}

\end{document}